\documentclass[prl,reprint,superscriptaddress]{revtex4-1}

\usepackage{graphicx}  
\usepackage{xcolor}
\usepackage{dcolumn}   
\usepackage{bm}       
\usepackage{amssymb}   
\usepackage{amsmath}
\usepackage{verbatim}
\usepackage[utf8]{inputenc}     
\usepackage{hyperref}
\usepackage{dcolumn}

\begin{document}

\title{Self-Interacting Dark Matter and the Origin of Ultra-Diffuse Galaxies NGC1052-DF2 and -DF4}

\author{Daneng Yang}
\email{yangdn@mail.tsinghua.edu.cn}
\affiliation{Department of Physics, Tsinghua University, Beijing 100084, China}
\author{Hai-Bo Yu}
\email{haiboyu@ucr.edu} 
\affiliation{Department of Physics and Astronomy, University of California, Riverside, California 92521, USA}
\author{Haipeng An}
\email{anhp@mail.tsinghua.edu.cn} 
\affiliation{Department of Physics, Tsinghua University, Beijing 100084, China}
\affiliation{Center for High Energy Physics, Tsinghua University, Beijing 100084, China}

\date{\today}

\begin{abstract}

Observations of ultra-diffuse galaxies NGC 1052-DF2 and -DF4 show they may contain little dark matter, challenging our understanding of galaxy formation. Using controlled N-body simulations, we explore the possibility that their properties can be reproduced through tidal stripping from the elliptical galaxy NGC 1052, in both cold dark matter (CDM) and self-interacting dark matter (SIDM) scenarios. To explain the dark matter deficiency, we find that a CDM halo must have a very low concentration so that it can lose sufficient inner mass in the tidal field. In contrast, SIDM favors a higher and more reasonable concentration as core formation enhances tidal mass loss. Final stellar distributions in our SIDM benchmarks are more diffuse than the CDM one, and hence the former are in better agreement with the data. We further show that a cored CDM halo model modified by strong baryonic feedback is unlikely to reproduce the observations. Our results indicate that SIDM is more favorable for the formation of dark-matter-deficient galaxies.

\end{abstract}

\maketitle

{\noindent\bf Introduction.} Dark matter plays a crucial role in galaxy formation and evolution~\cite{White:1977jf}. In the standard scenario, a luminous galaxy is hosted by a dark matter halo, which dominates the overall mass of the galactic system. However, using globular clusters as a tracer, the Dragonfly team found that ultra-diffuse galaxies NGC 1052-DF2 (DF2) and -DF4 (DF4) may contain little dark matter~\cite{vanDokkum:2018vup,vanDokkum:2018van,vanDokkum:2019va}. The ratio of their dark matter to stellar masses is $M_{\rm DM}/M_{\rm star}\sim1$, a factor of $300$ lower than expected from the canonical stellar-to-halo mass relation~\cite{Moster:2009fk,Moster:2012fk,Behroozi:2012iw}. There has been intensive debate in the literature about uncertainties that may affect the inference, including those associated with mass~\cite{Martin:2018ijt,Laporte:2018la,Hayashi:2018emo,Nusser:2018nu,Huo:2019hu} and distance~\cite{Trujillo:2018tr,Monelli:2019hfa} estimates. Further measurements~\cite{Danieli:2019zyi,Danieli:2019da} confirm the original findings, although more observations are needed to fully settle the debate.

The dark matter deficiency of these galaxies may be related to their environments. Both DF2 and DF4 are likely satellite galaxies of the massive elliptical galaxy NGC 1052~\cite{Danieli:2019da,vandokkum:2018va,Blakeslee:2018bl}, which is at a distance of $20~{\rm Mpc}$ from the earth. Cosmological hydrodynamical cold dark matter (CDM) simulations could produce dark-matter-deficient satellite galaxies~\cite{Jing:2018ji,Yu:2018wxs,Liao:2019li,Haslbauer:2019gzq} due to tidal stripping~\cite{Hayashi:2002qv,Penarrubia:2007zx,Penarrubia:2010jk,Errani:2015ep,vandenBosch:2017ynq,Sameie:2019zfo,Kahlhoefer:2019oyt}. However, it is challenging to find analogs in CDM simulations that could match properties of DF2 and DF4 even after taking into account the known uncertainties, as highlighted in~\cite{Haslbauer:2019cpl}. Ref.~\cite{Ogiya:2018jww} constructs a DF2-like system based on controlled N-body simulations. It argues that a cored dark matter halo is required, as a CDM cuspy halo does not lose sufficient mass in NGC 1052's tidal field. In addition, stars expand more significantly in a cored halo in response to tidal stripping~\cite{Carleton:2018ca}.

In this paper, we study realizations of DF2 and DF4-like galaxies in the self-interacting dark matter (SIDM) scenario~\cite{Spergel:1999mh,Kaplinghat:2015aga}; see~\cite{Tulin:2017ara} for a recent review. Dark matter self-interactions can thermalize the inner halo and naturally lead to a density core in the inner halo for low-surface brightness galaxies, see, e.g.,~\cite{Vogelsberger:2012ku,Rocha:2012jg,Zavala:2012us,Vogelsberger:2015gpr,Robertson:2016qef,Banerjee:2019bjp,Nadler:2020ulu}. Recent studies show that SIDM may provide a unified explanation for dark matter distributions in galactic systems over a wide mass range, including satellite galaxies in the Milky Way~\cite{Vogelsberger:2012ku,Valli:2017ktb,Sameie:2019zfo,Kahlhoefer:2019oyt}, spiral galaxies in the field~\cite{Kamada:2016euw,Creasey:2016jaq,Ren:2018jpt,Kaplinghat:2019dhn} and galaxy clusters~\cite{Kaplinghat:2015aga}. It is intriguing to see how the newly observed dark-matter-deficient galaxies can shed further light on the nature of dark matter.

Using controlled N-body simulations, we model the evolution of satellite galaxies in the tidal field of NGC 1052 and study their properties in both SIDM and CDM scenarios. After choosing a radial orbit to enhance tidal stripping, we impose observational constraints from the dark-matter-deficient galaxies and derive conditions on initial halo parameters for the satellites. We will show that SIDM is more likely to result in DF2 and DF4-like galaxies than CDM, in terms of reproducing their little dark matter content and diffuse stellar distributions. These results are insensitive to the initial halo mass. We further demonstrate that stellar particles can prevent a satellite halo from being disrupted in the tidal field, and a cored CDM halo, motivated by simulations with strong baryonic feedback, could be destroyed within $11~{\rm Gyr}$.

{\noindent\bf Simulation setup.} We model host galaxy NGC 1052 with a static spherical potential, including both dark matter halo and stellar components. Assuming a Navarro-Frenk-White (NFW) density profile~\cite{Navarro:1996gj} for the host halo, we fix the characteristic scale density and radius as $\rho_{\rm s}=1.6\times10^6~{\rm M_\odot/kpc^3}$ and $r_{\rm s}=80~{\rm kpc}$, similar to those used in~\cite{Ogiya:2018jww}. The total halo mass is $M_{200}=1.1\times10^{13}~{\rm M_\odot}$. The luminosity of the galaxy follows a 2D S{\'e}rsic profile~\cite{Sersic:1963se} with the index parameter $n=3.4$ and the effective radius $R_{\rm e}=2.06~{\rm kpc}$; the total stellar mass is $10^{11}~{\rm M_\odot}$~\cite{2017MNRAS.464.4611F}. In our simulations, we use a Hernquist profile $\rho_{\rm H}=\rho_{\rm h}/[r/r_{\rm h}(1+r/r_{\rm h})^3]$~\cite{Hernquist:1990be} to model the stellar distribution, where we take $\rho_{\rm h}=1.1\times10^{10}~{\rm M_\odot/kpc^3}$ and $r_{\rm h}=1.2~{\rm kpc}$ such that both Hernquist and 3D deprojected S{\'e}rsic profiles have the same total enclosed mass and half-mass radius.

For the satellite system, we consider a conservative scenario where tidal mass loss of stars is small, i.e., the ratio of initial to final stellar masses is ${\cal O}(1)$, so that the initial halo mass could be as low as possible, subject to the stellar-halo mass relation, which also holds in SIDM~\cite{Ren:2018jpt}. In this work, we choose the {initial} halo and stellar masses as $M_{200}=6.0\times 10^{10}~{\rm M_\odot}$ and $M_\star=3.2\times10^{8}~{\rm M_\odot}$, consistent with the median stellar-halo mass relation~\cite{Moster:2009fk,Moster:2012fk}. We assume that the halo follows an NFW profile and perform a coarse scan of the concentration parameter $c_{200}\equiv r_{200}/r_{\rm s}$, where $r_{200}$ is the halo's virial radius ($z=0$). We vary $c_{200}$ from $4$ to $10$, and find a value such that, after tidal evolution, the simulated satellites can match the observed dark matter content of DF2 and DF4. In practice, for a given halo, we convert its ($c_{200},~M_{200}$) to ($r_{\rm s},~\rho_{\rm s}$) for specifying its initial density profile in simulations. For the stellar component, we use a Plummer profile $\rho_{\rm P}=\rho_{\rm p}/[1+(r/r_{\rm p})^2]^{5/2}$~\cite{1911MNRAS..71..460P} to model its initial distribution, where $\rho_{\rm p}=5.8\times10^7~{\rm M_\odot/kpc^3}$ and $r_{\rm p}=1.1~{\rm kpc}$. Our stellar distribution is consistent with the size-mass relation found in galaxy surveys, see~\cite{Carleton:2018ca}.

We perform both SIDM and CDM simulations. For the former, we consider two values of the self-scattering cross section per mass, $\sigma/m=3~{\rm cm^2/g}$ (SIDM3) and $5~{\rm cm^2/g}$ (SIDM5), consistent with the ones used to explain dark matter distributions in field spiral and Milky Way satellite galaxies~\cite{Ren:2018jpt,Sameie:2019zfo}. We use the public code \texttt{GADGET-2}~\cite{Springel:2005mi,Springel:2000yr} to perform simulations. To model dark matter self-interactions, we have developed and implemented an SIDM module based on the method as in~\cite{Kochanek:2000pi}. We have checked our code for a test halo and found that the simulated density and velocity-dispersion profiles well agree with the results obtained using a semi-analytical model~\cite{Kaplinghat:2015aga}, which has been calibrated to other SIDM simulations, see~\cite{Ren:2018jpt}. For the satellite system, we use the code \texttt {SpherIC}~\cite{GarrisonKimmel:2013aq} to generate initial conditions. The mass of the simulation particles is $10^4~{\rm M_\odot}$ for both the halo and stellar components, and the softening length is $50~{\rm pc}$. We identify bound halo particles with the help of the code~\texttt{ROCKSTAR}~\cite{2013ApJ...762..109B}.

\begin{figure*}[t!]
	\includegraphics[scale=0.34]{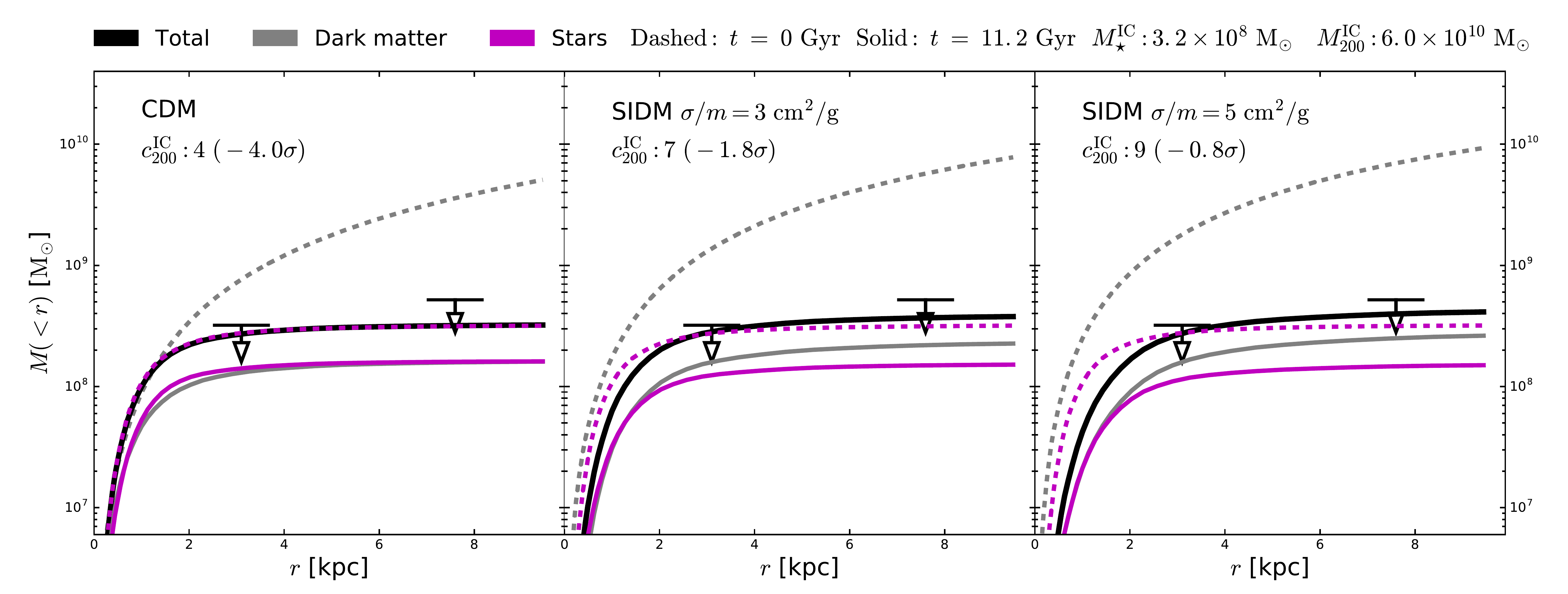}
\caption{Enclosed mass profiles before (dashed) and after (solid) tidal evolution for CDM, SIDM3 and SIDM5 benchmarks. The black arrows denote the upper limits on the total dynamical mass within $r=3.1~{\rm kpc}$ and $7.6~{\rm kpc}$ for the NGC 1052-DF2 galaxy at $90\%$ CL~\cite{vanDokkum:2018vup,vanDokkum:2018van}. As a reference, values in the parentheses indicate deviations of the initial halo concentrations from the median of the concentration-mass relation ($z=0$) predicted in cosmological simulations~\cite{Dutton:2014xda}.  }
\label{fig:mass}
\end{figure*}

{\noindent\bf Orbital parameters.} The simulated satellites are initially placed at the apocenter and they have a tangential velocity of $27~{\rm km/s}$. The apocenter is $380~{\rm kpc}$ away from the center of the host. We find that the orbital period is about $2.5~{\rm Gyr}$, the pericenter is $9~{\rm kpc}$, and the velocity there is $740~{\rm km/s}$. The corresponding orbital energy and circularity parameters are $x_{\rm c} = 0.49$ and $\eta=0.13$, respectively, which are on lower tails of cosmological distributions~\cite{vandenBosch:2017ynq}. We have chosen a radial orbit to enhance tidal stripping of the halo mass.

We determine a timescale for the final snapshot from the following consideration. DF2 has a projected distance of $80~{\rm kpc}$ from NGC 1052 and a relative velocity of $293~{\rm km/s}$ along the line-of-sight direction~\cite{vanDokkum:2018vup}. Suppose the angle between the host--satellite plane and the line-of-sight direction is $\theta$ and the orbit is nearly radial, as in our setup, we have the relations, $d\sin\theta\approx80~{\rm kpc}$ and $v\cos\theta\approx293~{\rm km/s}$. For $t\approx9.6~{\rm Gyr}$ and $11.2~{\rm Gyr}$, corresponding to the moments right before and after passing the apocenter for the fourth time, respectively, our simulated satellites satisfy condition $\left({80~{\rm kpc}}/{d}\right)^2+\left({293~{\rm km/s}}/{v}\right)^2\approx1$. We will show results with $t=11.2~{\rm Gyr}$, corresponding to $4$ orbits in total.

We make additional checks to further validate our simulations. Using a low-resolution simulation that treats the host halo with live particles, we find that dynamical friction could shorten the orbital period by $10\%$ at late stages, which is negligible for the purpose of this work. For our system, the evaporation effect could be important if the cross section between dark matter in the host and the one in the satellite is $\sigma/m\gtrsim2.5~{\rm cm^2/g}$. Consider a realistic SIDM model in~\cite{Kaplinghat:2015aga}, where $\sigma/m$ is velocity dependent and diminishing towards cluster scales, we estimate $\sigma/m\sim0.3~{\rm cm^2/g}$ when the satellite passes the pericenter at a velocity of $v=740~{\rm km/s}$, and thus evaporation mass loss of the satellite halo is negligible. Note that $\sigma/m$ can be much larger in the satellite halo as the dark matter velocity there is much lower, $\sim40~{\rm km/s}$. 

We also simulate CDM and SIDM ($\sigma/m=3~{\rm cm^2/g}$) halo density profiles for NGC 1052 in the presence of the stellar potential, and find that they behave similarly, i.e., both of them are steeper than the assumed NFW profile towards the center. However, since stars dominate the host's inner potential, the difference in the halo density profile of the host has a small effect. Thus our approach with the static potential is well justified. Lastly, using the semi-analytic method in~\cite{vandenBosch:2018tyt}, we have confirmed that resolution of our benchmark simulations is high enough to avoid numerical artifacts concerning tidal disruption of substructure in N-body simulations~\cite{vandenBosch:2017ynq,vandenBosch:2016hjf}.

{\noindent\bf Mass profiles.} Our simulations search for upper limits of $c_{200}$ such that the simulated satellites can broadly match the observations. We find three benchmark cases, $c_{200}=4,~7,$ and $9$ for CDM, SIDM3 and SIDM5, respectively. Fig.~\ref{fig:mass} shows their enclosed dark matter (gray) and stellar (magenta) masses vs. radius at $t=0~{\rm Gyr}$ (dashed) and $11.2~{\rm Gyr}$ (solid), as well as the final total mass profiles (solid black). For all the benchmarks, the simulated halos experience significant mass loss in the tidal field of NGC 1052, and the bound masses approach $M_{\rm DM}\approx2\times10^8~{\rm M_\odot}$ at $t=11.2~{\rm Gyr}$, reduced by a factor of $300$ compared to the initial value, $6\times10^{10}~{\rm M_\odot}$. Tidal mass loss of the stars is much more mild, resulting in a total stellar mass of $M_{\rm star}\approx 1.5\times10^8~{\rm M_\odot}$. Thus the mass ratio is $M_{\rm DM}/M_{\rm star}\sim1$ after tidal evolution.

Observationally, Ref.~\cite{vanDokkum:2018vup} reports a $90\%$ CL upper limit of $\sigma_{\rm intr}<10.5~{\rm km/s}$ for the velocity dispersion of $10$ globular clusters of DF2, corresponding to a total dynamical mass of $M_{\rm dyn}\lesssim3.4\times10^{8}~{\rm M_\odot}$ within a radius of $7.6~{\rm kpc}$; see also~\cite{Emsellem:2018em,Fensch:2018fe} using other tracers. Ref.~\cite{vanDokkum:2018van} revises it to be $\sigma_{\rm intr}<12.4~{\rm km/s}$, and we obtain $M_{\rm dyn}\lesssim4.7\times10^{8}~{\rm M_\odot}$ accordingly ($M_{\rm dyn}\propto\sigma^2_{\rm intr}$~\cite{Watkins:2010fe}). DF2 has a total stellar mass of $M_{\rm star}=2\times10^{8}~{\rm M_\odot}$~\cite{vanDokkum:2018vup}. Further measurements using stellar spectroscopy show $M_{\rm star}=(1.0\pm0.2)\times10^8~{\rm M_\odot}$ and $M_{\rm dyn}=(1.3\pm0.8)\times10^8~{\rm M_\odot}$ within DF2's 3D half-light radius $R_{1/2}=2.7~{\rm kpc}$~\cite{Danieli:2019zyi}. For DF4, $M_{\rm star}=(1.5\pm0.4)\times10^8~{\rm M_\odot}$ and $M_{\rm dyn}=0.4^{+1.2}_{-0.3}\times10^8~{\rm M_\odot}$ within $7~{\rm kpc}$~\cite{vanDokkum:2019va}. We see our three benchmarks well reproduce low dark matter content of DF2 and DF4. For reference, we display the upper limits of $M_{\rm dyn}$ for DF2 in Fig.~\ref{fig:mass}, $3.2\times10^8~{\rm M_\odot}$ and $4.7\times10^8~{\rm M_\odot}$ within $3.1~{\rm kpc}$ and $7.6~{\rm kpc}$, respectively.

{\noindent\bf Halo concentration.} Using tailored simulations, we have shown that the tidal interactions can cause low dark matter content of DF2 and DF4 in both CDM and SIDM scenarios. However, the benchmark halos shown in Fig.~\ref{fig:mass} have different concentrations. The CDM one is the lowest $c_{200}=4$, corresponding to $4\sigma$ below the median of the concentration-mass relation~\cite{Dutton:2014xda} ($z=0$). In contrast, SIDM favors a higher concentration. SIDM3 has $c_{200}=7$, $1.8\sigma$ below the median, and SIDM5 has $c_{200}=9$, $0.8\sigma$ below. 

We see that the dark-matter-deficient galaxies are more likely realized in SIDM than in CDM. Since the inner density cusp in a CDM halo is resilient to tidal stripping, a low concentration is required. In contrast, dark matter self-interactions can thermalize the inner halo and push dark matter from inner to outer regions, lowering the inner gravitation potential. Thus SIDM satellite halos are more prone to tidal stripping (if there is no core collapse; see~\cite{Sameie:2019zfo,Kahlhoefer:2019oyt}), and a higher and more reasonable $c_{200}$ value can match the observations.

To further test our findings, we have performed a series of additional CDM simulations, where we keep the same initial stellar mass, but varying the halo mass within $\pm1\sigma$ deviations from the median stellar-to-halo mass relation~\cite{Moster:2012fk}. Even with the flexibility, the CDM halo concentration needs to be $\sim4\sigma$ below the median~\cite{Dutton:2014xda} to achieve $M_{\rm DM}/M_{\rm star}\sim1$. Thus the CDM model is strongly disfavored, regardless of the initial halo mass. We also have checked that if the SIDM halos evolve for $2~{\rm Gyr}$ in isolation before being subject to the tidal field, an even higher concentration is allowed. Thus our results are robust.

\begin{figure*}[t!]
	\includegraphics[scale=0.40]{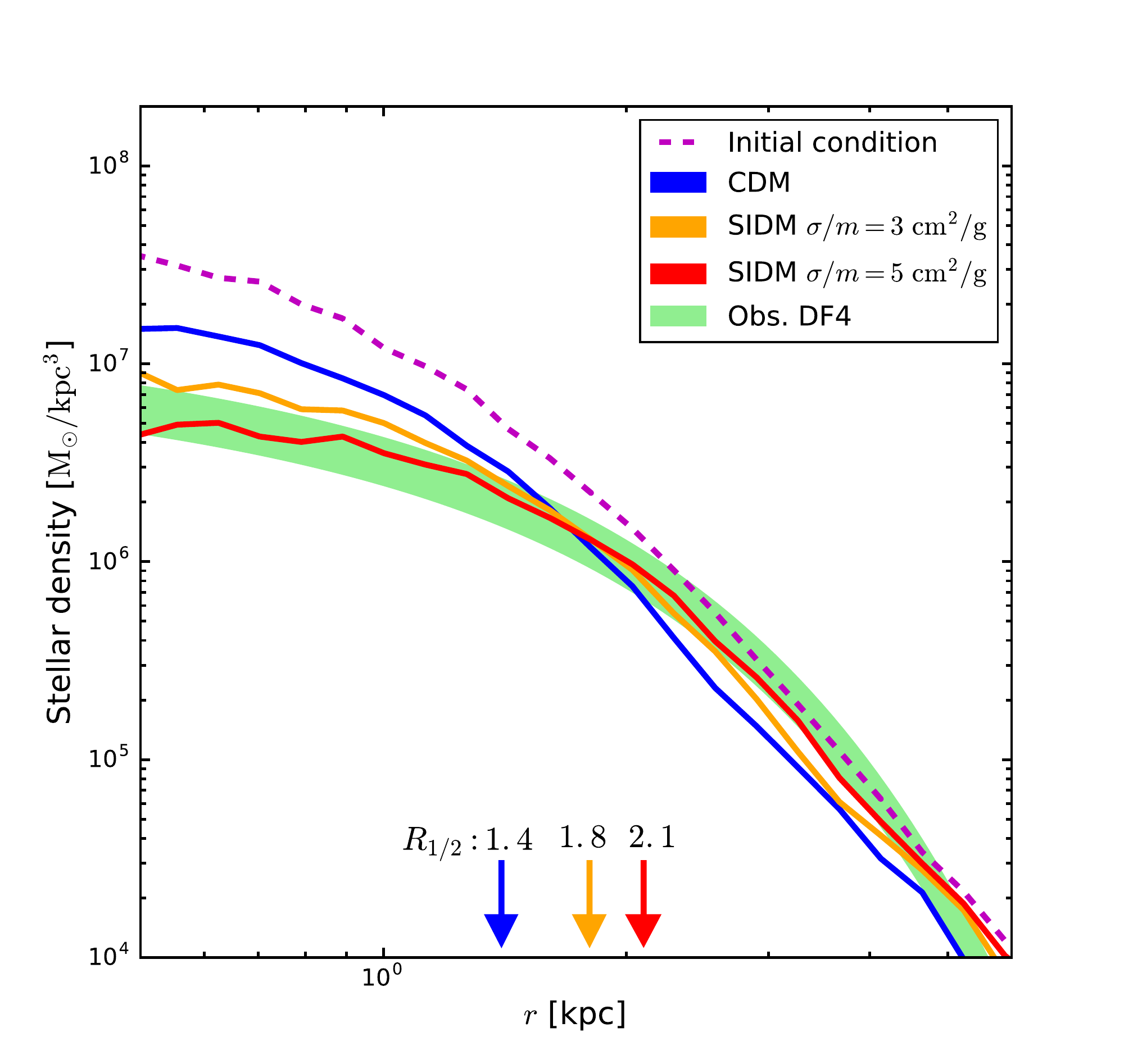}\;\;\;\;
				\includegraphics[scale=0.40]{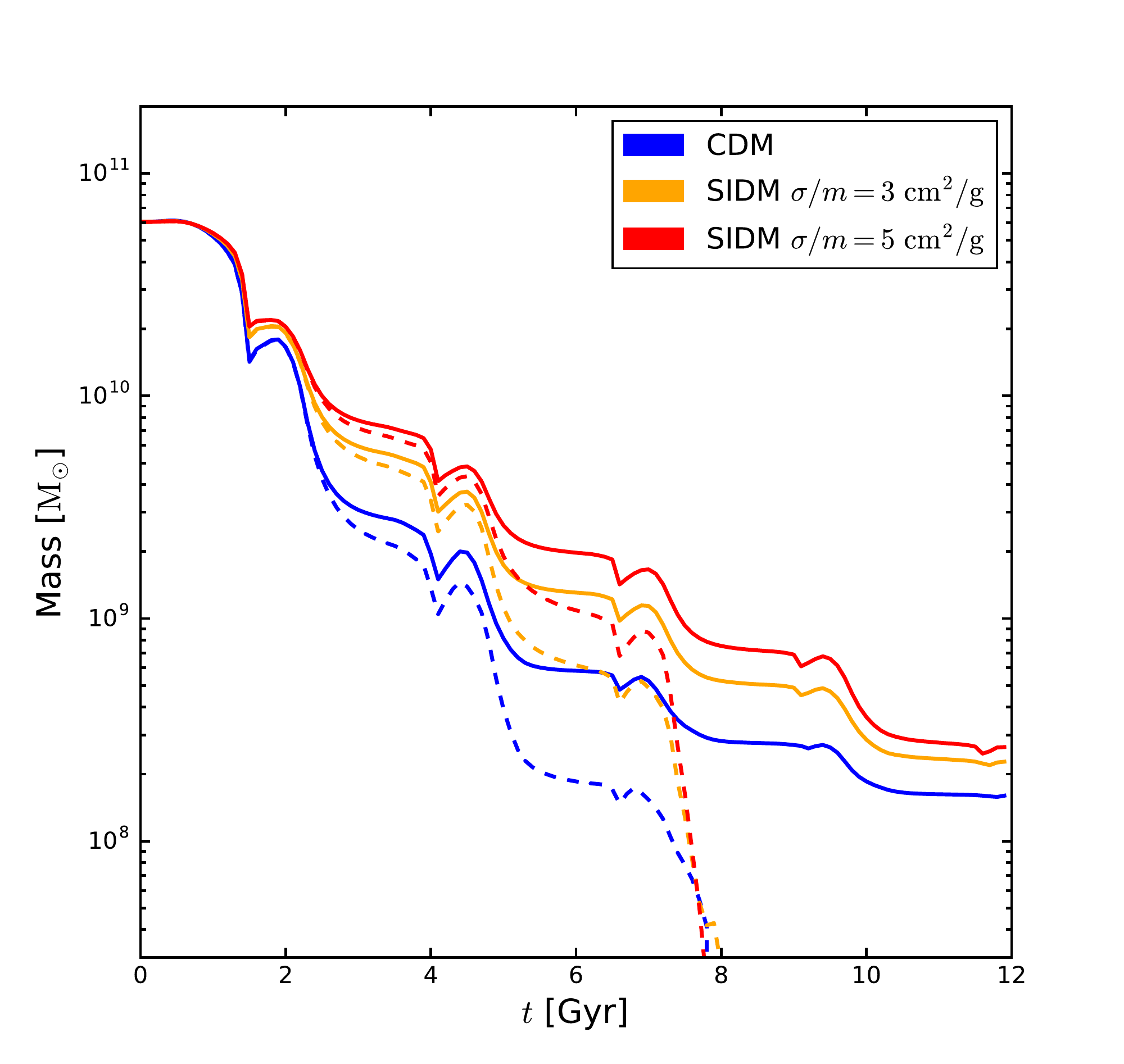}
\caption{{\it Left:} final stellar density profiles for the CDM, SIDM3 and SIDM5 benchmarks (solid), as well as the initial condition (dashed). The green band denotes the measured stellar density of NGC 1052-DF4~\cite{vanDokkum:2019va}, and the arrows denote the half-mass radii of the stellar components from the simulations. {\it Right:} evolution of the bound halo masses for the benchmarks (solid). For comparison, we also show results without including stellar particles in simulations (dashed).}

\label{fig:stellar}
\end{figure*}

{\noindent\bf Stellar distributions.} Fig.~\ref{fig:stellar} (left) shows stellar density profiles at $t=0~{\rm Gyr}$ and $11.2~{\rm Gyr}$ for the benchmarks. From our simulations, we find that final half-mass radii are $R_{1/2}=1.4~{\rm kpc},~1.8~{\rm kpc}$ and $2.1~{\rm kpc}$ for CDM, SIDM3 and SIDM5 benchmarks, respectively. We use a 3D deprojected S\'esic profile to fit the simulated stellar distributions within $5~{\rm kpc}$ and find good agreement. The inferred S\'esic indices, characterizing the stellar concentration, are $n=1.0,~0.81$ and $0.67$ for CDM, SIDM3 and SIDM5, respectively; the associated effective radii $R_{\rm e}$ are consistent with the half-mass radii from the simulations ($R_{1/2}\approx4/3R_{\rm e}$~\cite{Wolf:2009tu}). 

As $\sigma/m$ increases, the stellar distributions become more diffuse and the baryon concentration decreases. This is because SIDM core formation leads to a shallow gravitational potential and the stars expand more significantly through tidal stripping; see also~\cite{Carleton:2018ca}. The measured half-light radii and S\'esic indices are $R_{1/2}=2.7~{\rm kpc}$ and $n=0.6$ for DF2~\cite{vanDokkum:2018vup}, and $R_{1/2}=2.0~{\rm kpc}$ and $n=0.79$ for DF4~\cite{vanDokkum:2019va}. In Fig.~\ref{fig:stellar} (left), we show the measured stellar density of DF4 for comparison. Our SIDM benchmarks agree better with the observed stellar distributions than the CDM one, although all three cases have $M_{\rm DM}/M_{\rm star}\sim1$ after evolution, as shown in Fig.~\ref{fig:mass}.

{\noindent\bf Tidal evolution.} Fig.~\ref{fig:stellar} (right) shows the bound halo masses vs. time for the benchmarks (solid). There is a trend between the concentration of the benchmark models and the rate of mass loss, i.e., the SIDM5 halo ($c_{200}=9$) has much less rapid mass loss than the CDM halo ($c_{200}=4$). We also display the evolution of the bound halo masses {\it without} including stellar particles in the simulations (dashed). All of the halos are destroyed right after their third pericenter passages at $t\approx7.5~{\rm Gyr}$. Thus the stars are crucial for preventing the simulated halos from being disrupted in the tidal field. Even though the SIDM5 halo has a high concentration, dark matter self-interactions make it vulnerable to tidal disruption in the absence of the stars. For comparison, we perform CDM runs of the initial SIDM5 halo without stellar particles. The halo survives from tidal disruption and its final bound mass is $2.6\times10^8~{\rm M_\odot}$.

Our results may have implications for understanding the population of dark-matter-deficient galaxies. If CDM satellite halos host DF2 and DF4 galaxies, they need to be on the very low end of the concentration distribution. Thus a large number of satellite halos with higher concentrations would populate the NGC 1052 group. They would have deep gravitational potentials to collect gas to form stars and prevent them from tidal disruption. It seems odd that we have observed only two that contain little dark matter. On the other hand, SIDM satellite halos have higher concentrations towards the median, and they are more prone to disruption. Thus it is natural to expect that luminous satellite galaxies in the NGC 1052 group are rare in SIDM. It would be interesting to test this scenario with cosmological simulations and future observations.

{\noindent\bf Discussion.} Dark matter self-interaction is not the only mechanism that can create density cores. Hydrodynamical simulations show that strong baryonic feedback may create cores in CDM halos~\cite{Navarro:1996bv,Governato:2009bg,2013MNRAS.429.3068T,DiCintio:2013qxa,Chan:2015tna,Read:2015sta,2016MNRAS.456.3542T,Fitts:2016usl,Hopkins:2017ycn,2018MNRAS.473.4392S,DiCintio:2019bxz,Lazar:2020pjs}. For DF2-like systems under consideration, the ratio of initial stellar-to-halo masses is $\sim5\times10^{-3}$, at which the feedback has the strongest impact and the core size is the largest~\cite{DiCintio:2013qxa,2013MNRAS.429.3068T,Lazar:2020pjs}. Ref.~\cite{Ogiya:2018jww} considers a cored CDM halo motivated by those simulations and finds it could reproduce DF2 observations after tidal evolution, although the halo has a median concentration. It also assumes a steep initial density profile for stars, which dominate in mass for $r\lesssim1.5~{\rm kpc}$.

However, CDM simulations that create dark matter cores would also produce diffuse stellar distributions, as argued in~\cite{Kaplinghat:2019dhn}. Consider a simulated galaxy from recent FIRE-2 simulations (m10xg)~\cite{Lazar:2020pjs}, which has a dark matter core due to strong feedback. Its halo mass is $6.2\times10^{10}~{\rm M_\odot}$ and stellar mass $4.6\times10^8~{\rm M_\odot}$, similar to our benchmarks. However, its stellar Plummer radius is $3.1~{\rm kpc}$, a factor of $3$ larger than ours, and its central stellar density is $15$ times lower. We evolve a m10xg-like system in the tidal field of NGC 1052. It is destroyed at $t\sim7~{\rm Gyr}$ when using our orbital parameters, and $11~{\rm Gyr}$ using those in~\cite{Ogiya:2018jww}. The ratio $M_{\rm DM}/M_{\rm star}$ has a minimum of $\sim5$ right before tidal disruption, at which the halo mass is close to $10^9~{\rm M_\odot}$. Thus the cored CDM halo modified by strong feedback is unlikely to reproduce DF2- and DF4-like systems. Note that our initial stellar distribution is broadly consistent with simulations with weak baryonic feedback as in~\cite{Vogelsberger:2014pda}.

{\noindent\bf Conclusions.} We have studied realizations of dark-matter-deficient galaxies through tidal stripping. Both CDM and SIDM halos can lose the majority of their mass in the NGC 1052's tidal field, drastically increasing the ratio of stellar-to-halo masses in accord with the observations. In CDM, halo concentration needs to be low to explain the dark matter deficiency. In contrast, an SIDM halo can have a higher and more reasonable concentration, as collisional thermalization leads to core formation, boosting tidal mass loss. Our SIDM benchmarks also predict more diffuse stellar distributions, resulting in better agreement with measurements. Thus the newly observed DF2 and DF4 are more naturally realized in SIDM than in CDM scenarios. There are interesting topics we could explore further, such as correlations between orbital and halo parameters of the satellite, and formation of NGC 1052-like systems in the cosmological setup. Recent observations of dark-matter-deficient galaxies in the field~\cite{Guo:2019wgb,Pina:2019rer,Pina:2020krc} may help test the nature of dark matter from another angle, as the tidal effects are absent in these systems.

 \vspace{0.3in}
  
\begin{acknowledgments}
We would like to thank Ran Huo and Go Ogiya for useful discussions. HBY was supported by the U.S. Department of Energy under Grant No. de-sc0008541 and in part by the U.S. National Science Foundation under Grant No. NSF PHY-1748958 through the ``From Inflation to the Hot Big Bang" KITP program. HA was supported by NSFC under Grant No. 11975134, the National Key Research and Development Program of China under Grant No. 2017YFA0402204 and Tsinghua University Initiative Scientific Research Program.

\end{acknowledgments}

\bibliography{reference}

\end{document}